\documentclass[aps,prx,twocolumn,amsmath,nofootinbib,longbibliography,amssymb,superscriptaddress,10pt,floatfix]{revtex4-2}
\usepackage[colorlinks, linkcolor=Blue , citecolor= Blue,urlcolor = NavyBlue, breaklinks=true]{hyperref}
\usepackage[english]{babel}
\usepackage[svgnames]{xcolor} 
\usepackage{verbatim}
\usepackage{braket}
\usepackage{esint}
\usepackage{mathtools,bm,bbm}
\usepackage{multirow}

\usepackage{tikz-qtree}
\usepackage{tikz}

\begin{document}
\title{
Incommensurate inter-valley coherent states
\texorpdfstring{\\}{ }
in ABC graphene: collective modes and superconductivity
}
\author{Yaar Vituri}\thanks{These authors contributed equally to this work.}
\affiliation{Department of Condensed Matter Physics, Weizmann Institute of Science, Rehovot 76100, Israel}
\author{Jiewen Xiao}\thanks{These authors contributed equally to this work.}
\affiliation{Department of Condensed Matter Physics, Weizmann Institute of Science, Rehovot 76100, Israel}
\author{Keshav Pareek}
\affiliation{Department of Condensed Matter Physics, Weizmann Institute of Science, Rehovot 76100, Israel}
\author{Tobias Holder}
\affiliation{School of Physics and Astronomy, Tel Aviv University, Tel Aviv, Israel}
\author{Erez Berg}
\affiliation{Department of Condensed Matter Physics, Weizmann Institute of Science, Rehovot 76100, Israel}

\date{\today}

\begin{abstract}

    Recent experiments in ABC trilayer graphene detected superconductivity on the border of a phase transition to a symmetry-broken phase. In this work, we use unrestricted Hartree-Fock to study the nature of this phase. We find a close competition between two incommensurate inter-valley coherent (IVC) phases: an IVC crystal where the ordering occurs at multiple wavevectors, and an IVC spiral with a single ordering wavevector. Focusing on one of the regimes where superconductivity is observed experimentally, we find a continuous (or very weakly first order) transition between a half metallic phase to an IVC crystal, followed by a first-order transition into an IVC spiral. Using time-dependent Hartree-Fock, we study the collective mode spectrum in the half-metalic phase. We find a soft inter-valley mode that can mediate superconductivity in a narrow sliver of density near the continuous transition, with a $T_c$ that can reach a few hundreds of mK and a sign-changing s-wave order parameter. The spin stiffness in the half metal phase is found to be surprisingly low, of the order of a few degrees Kelvin.
    
\end{abstract}
\maketitle

{\it Introduction --} ABC stacked rhombohedral trilayer graphene has recently emerged as highly tunable platform for studying correlated electron phenomena. 
It shows a rich phase diagram, including symmetry broken metallic and superconducting phases~\cite{zhou2021half,zhou2021superconductivity,arp2023intervalley,winterer2023ferroelectric}. 
Similar phenomena are also observed in Bernal bilayer graphene~\cite{zhou2022isospin,seiler2022quantum,de2022cascade,zhang2023enhanced,seiler2024layer,li2024tunable,pantaleon2023superconductivity}. 
These observations have been explained in the language of flavor cascades~\cite{zhou2021half,zondiner2020cascade,wong2020cascade} and magnetic ordering~\cite{zhou2021half,arp2023intervalley,chatterjee2022inter,you2022kohn,szabo2022metals}.
Much of the phase diagram, including the pattern of symmetry breaking in many of the ordered phases, can be captured within self-consistent Hartree Fock (SCHF) calculations~\cite{zhou2021half, chatterjee2022inter, koh2024correlated, arp2023intervalley,Huang2023,wang2024electrical}. 

Crucially, superconductivity is often found on the border of a phase whose precise nature has not yet been identified.
Several mechanisms for superconductivity have been suggested, including a conventional phonon-mediated pairing~\cite{chou2021acoustic,chou2022acoustic,vinas2024phonon} and various unconventional mechanisms driven by electronic fluctuations~\cite{szabo2022metals,lu2022correlated,qin2023functional,dai2023quantum,dong2023signatures,ghazaryan2021unconventional,chatterjee2022inter,you2022kohn,dong2021superconductivity,cea2022superconductivity,jimeno2023superconductivity}. 
To assess the feasibility of a purely electronic mechanism, it is crucial to identify the nature of the electronic fluctuations and to quantify their strength. 

In this Letter, we aim to resolve the nature of the phases that border the superconducting regions in ABC graphene, and analyze collective modes and fluctuations near the transition into these phases. 
We propose that these phases are intervalley coherent phases whose ordering wavevector is slightly incommensurate with the graphene lattice.

Specifically, within unrestricted SCHF (allowing for incomensurate density wave order), we find two closely competing phases.
The first, denoted as \emph{IVC crystal}, hosts multiple finite momentum IVC order parameters and breaks translation invariance while preserving $C_3$-rotation symmetry. This phase was anticipated in previous works~\cite{you2022kohn,blinov2023partial}.  
The second is an \emph{IVC spiral} with a single order parameter with an incommensurate momentum, analogous to the one found
in magic-angle twisted bilayer and trilayer graphene~\cite{kwan2021kekule,nuckolls2023quantum,kim2023imaging}. 

Focusing on the superconducting region that occurs within the spin polarized half-metallic phase in ABC graphene (denoted SC2 in Ref. \cite{zhou2021superconductivity}), we find a sequence of transitions in the normal state: upon decreasing the hole density, the half-metal undergoes a second-order (or very weakly first-order) transition into a narrow spin-polarized IVC crystal phase, followed by a first-order transition into a spin-polarized IVC spiral phase. 
Using time-dependent Hartree-Fock (TDHF), we calculate the collective mode spectrum near the transition from the half-metal to the IVC crystal, and find soft IVC collective modes associated with a strongly enhancement of the intervalley susceptibility.
We extract the effective interaction induced by the coupling of the electrons to this mode, and solve the resulting linearized gap equations, including the bare Coulomb repulsion.  
We find that the leading instability is towards a sign-changing $s$-wave order parameter, whose $T_c$ can reach a few hundred mK in the vicinity of the transition.

\begin{figure}
    \centering
    \includegraphics[width=\linewidth]{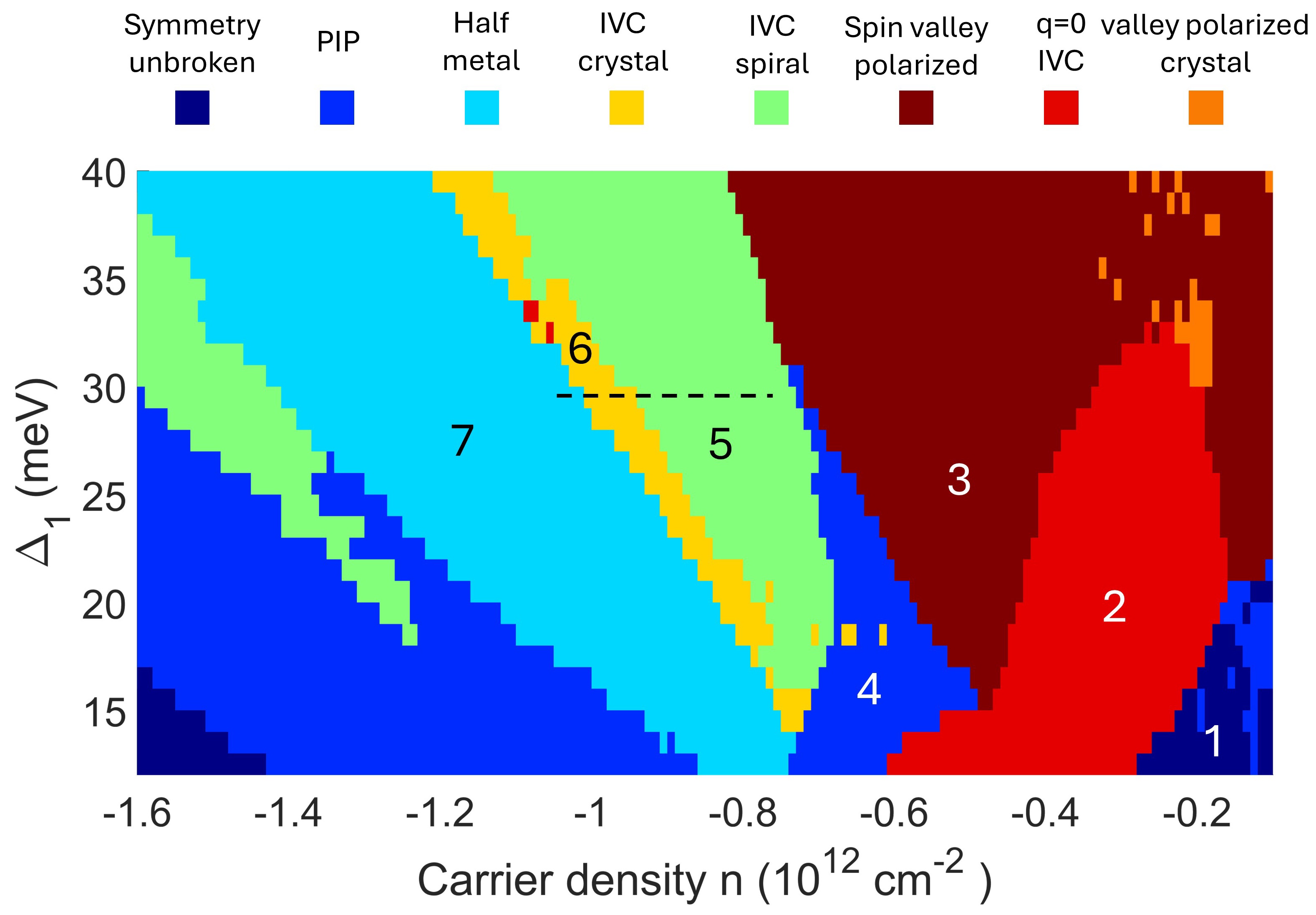}
    \caption{
     The phase map as a function of hole-doped carrier density n and layer potential $\Delta_1$ (meV) from self-consistent Hartree-Fock calculations. In the phase map, we identify the following phases: spin-valley polarized phase, commensurate IVC phase, IVC spiral phase, IVC crystal phase, partially isospin polarized (PIP) phase, and symmetry unbroken phase.
    }
    \label{fig1}
\end{figure}

{\it Model and Phase diagram --}
We use a tight binding model with long range Coulomb interactions to model the ABC trilayer graphene. 
The tight binding parameters are similar to those of Ref.~\cite{zhou2021half}, with modified interlayer hopping parameters to better fit the experimental data (See Table~\ref{tab:tight_binding} in the Supplementary Material (SM)~\cite{SM}). The perpendicular displacement field opens a gap between the conduction and valence band, allowing us to project the Hamiltonian to a single active band. 

We use a dual gated Coulomb interaction with an effective dielectric constant $\varepsilon_r=8$, representing the screening effect from the ABC graphene remote bands and the encapsulating boron nitride. 
To capture the screening by the conduction electrons at finite densities,
we include a phenomenological Thomas-Fermi screening~\cite{chatterjee2022inter} with $q_{TF} = 0.04/a_0$,
where $a_0$ is the lattice constant of graphene.
The Thomas Fermi screening length is of order $\lambda_{\mathrm{F}}$, consistent with a dimensionless interaction strength $r_s$ of order one. 
For simplicity, we neglect large momentum transfer Hund-type couplings, which are expected to be significantly weaker than the long-range density-density interactions.

Fig.~\ref{fig1} shows the zero temperature phase diagram as a function of carrier density $n$ and the inter-layer potential $\Delta_1$ (proportional to the perpendicular displacement field in the experiment). 
In the low density and displacement field region, the ground state does not break any symmetry (region 1). 
Increasing the carrier density, one finds a commensurate IVC phase (region 2), followed by a spin and valley polarized phase (region 3) and partially isospin polarized phase (PIP, region 4). 
This sequence of transitions closely resembles the experimental phase diagram~\cite{arp2023intervalley}.
Near this PIP phase, we find an IVC spiral and an IVC crystal phase (regions 5 and 6, respectively), which are the focus of this work.

{\it Incommensurate IVC states --} 
The IVC order parameter $\phi_{\bm{q}}$ is defined as the expectation value of the IVC operator $\hat{\phi}_{\bm{q}}=\frac{1}{A}\sum_{\bm{k}}c_{+,\bm{k}+\bm{q}}^\dagger c_{-,\bm{k}}$, where $c^\dagger_{\pm,\bm{k}}$ creates an electron in valley $\pm$, and $A$ is the system's area.
Here, for simplicity, we drop the spin indices (regions 5, 6, 7 above are all spin polarized).
In the full graphene Brillouin zone, this translates to a charge density wave at an incommensurate momentum $\bm{K}+\bm{q}$. 
The $\bm{q}=0$ case correspond to a commensurate density wave that triples the unit cell in real space. 

Symmetry-wise, an IVC with $\bm{q}\ne 0$ 
is distinct from the $\bm{q}=0$ one because these states break different symmetries. In the IVC spiral phase, where intervalley order appears at a single ordering vector $\bm{q}$, $C_3$ symmetry is spontaneously broken. Within the continuum model, continuous translational symmetry is preserved and generated by the modified translation operator $\hat{T}^{\prime}_{\bm{r}}=\hat{T}_{\bm{r}}e^{i(\bm{q}\cdot \bm{r}) \tau_z/2}$~\cite{kwan2021kekule}, where $\hat{T}_{\bm{r}}$ is the translation operator by $\bm{r}$ and $\tau_z$ is a Pauli matrix in valley space.

\begin{figure}
    \centering
    \includegraphics[width=\linewidth]{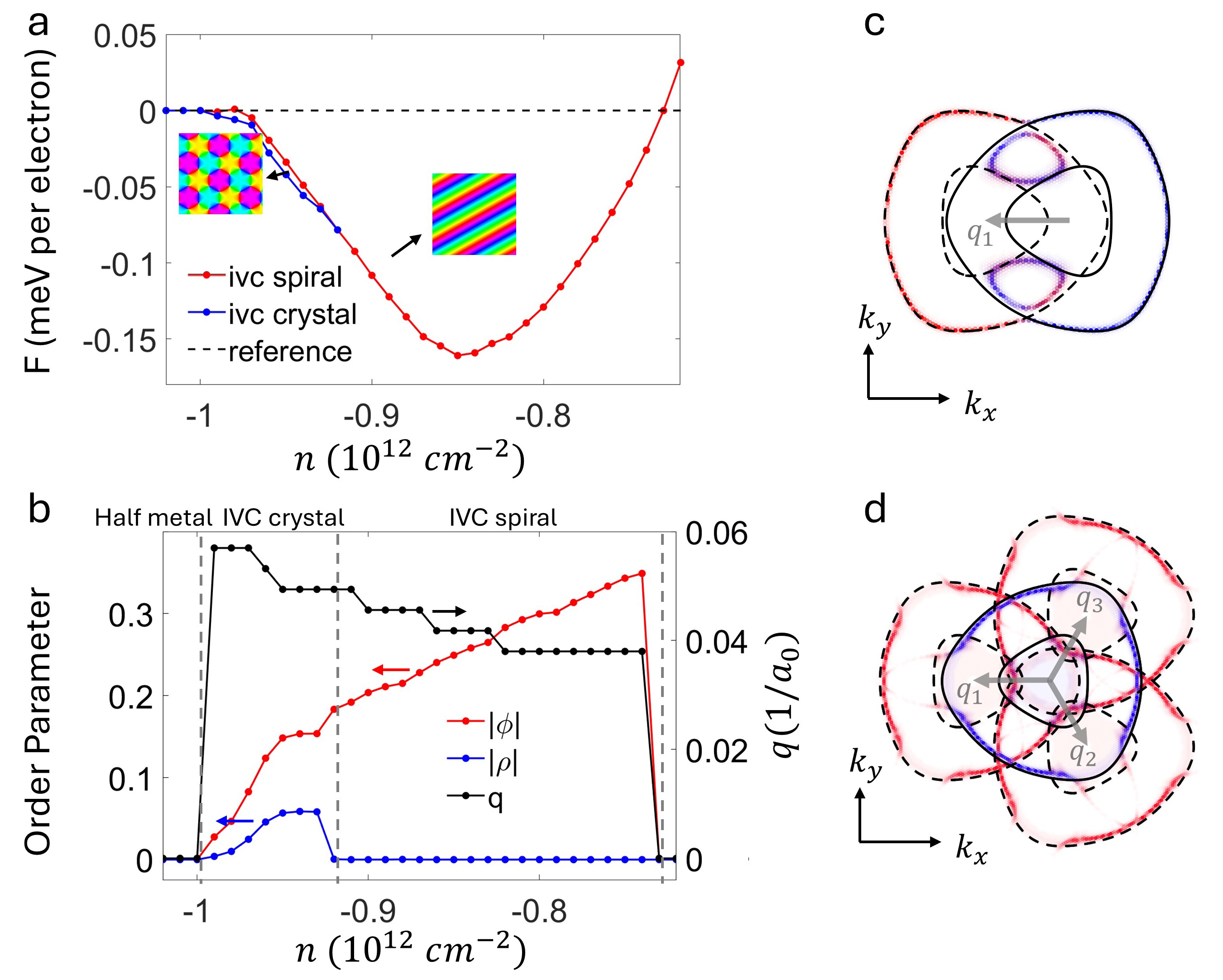}
    \caption{
    a) Ground state energy versus density at $\Delta_1=28$ meV for IVC spiral, IVC crystal and the spin-valley diagonal phase (reference line). This parameter range is marked by the black dashed line in Fig. \ref{fig1}. Insets show the complex phase of the IVC order parameter in real space for both the crystal and the spiral phases. 
    b) The corresponding IVC ($\phi$) and CDW ($\rho$) order parameters as function of density. As discussed in the text, the CDW order is a direct consequence of the IVC crystal phase. The black line shows the ordering vector $q$ of the IVC. 
    c) Spectral function at zero frequency of the IVC spiral phase. A broadening of $0.3 \text{ meV}$ was used. The color represents the valley polarization (red for $K$ valley and blue for $K'$ valley). The dashed (solid) line represents the non-interacting FS of the $K$ ($K'$) valley. The two valleys are plotted shifted by the IVC ordering vector $\bm{q}_1$ (in grey). 
    d) Same as c) for IVC crystal phase, where the $K$ valley 
    shifted by either of the three ordering vectors $\bm{q}_i$ (in grey) with respect to the $K'$ valley. Here, the outer FS is similar to the non-interacting one, except for hybridization at three distinct points. The inner one is largely gapped through hybridization. 
    }
    \label{fig2}
\end{figure}

\begin{figure*}
    \centering
    \includegraphics[width=\linewidth]{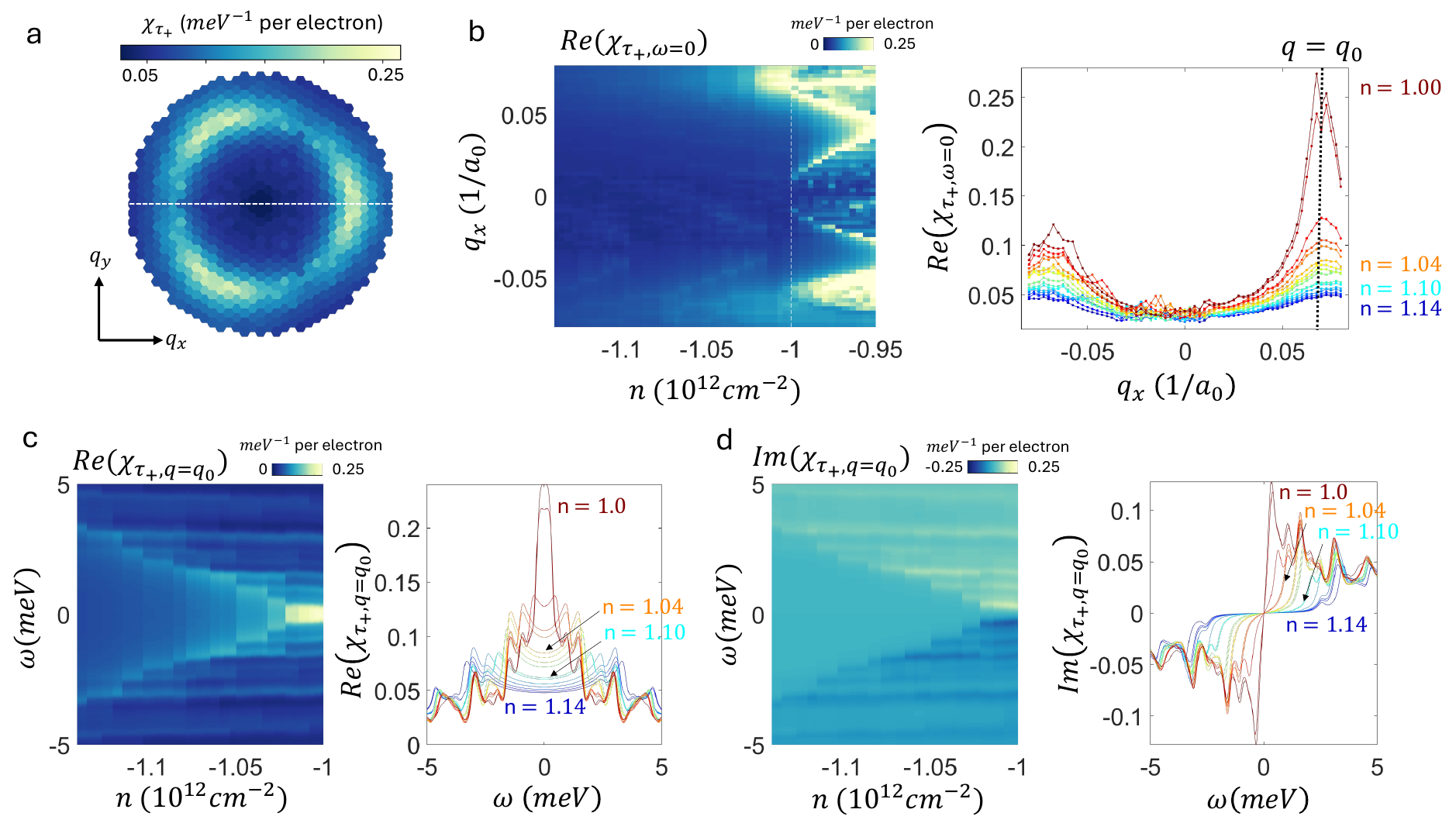}
    \caption{a) Static intervalley susceptibility $\chi_{\tau_+,\omega=0}$ in the half-metal phase as a function of $\bm{q}$ at carrier density $-1.0 \times 10^{12} \text{cm}^{-2}$. 
    b) Intervalley susceptibility $\chi_{\tau_+}$ in the half-metal phase as a function of $q_x$ (while dashed line in Fig. \ref{fig3}a) and carrier density. 
    The dashed line marks the phase boundary between half metal and IVC crystal. The right panel shows $\chi_{\tau_+}$ at $\omega = 0$ as a function $q_x$ for a density range from $-1.1 \times 10^{12} \text{cm}^{-2}$ (deep in the half metal phase region) to $-1.0 \times 10^{12} \text{cm}^{-2}$ (near the phase transition). 
    c) and d), the real and imaginary part of $\chi_{\tau_+}$ at the momentum of the peak in $\chi_{\tau_+,\omega=0}(\bm{q})$, 
 $\bm{q} = \bm{q}_0$, as a function of $\omega$ and carrier density. The right panel plots line cuts as a function of $\omega$ for the same density range in Fig. \ref{fig3}b.}
    \label{fig3}
\end{figure*}

In the IVC crystal phase $C_3$-symmetry is preserved, while continuous translation symmetry is broken to a discrete subgroup. This phase has three order parameters $\phi_{\bm{q}_i}$ of equal magnitude with  ordering vectors $\bm{q}_i$ related by $C_3$ according to $\bm{q}_3=R_{2\pi/3}\cdot \bm{q}_2=R_{4\pi/3}\cdot \bm{q}_1$, where $R_{\theta}$ is a rotation matrix by $\theta$. 
The discrete translation symmetry is generated by $\hat{T}^{\prime}_{i} =\hat{T}_{\bm{R}_i} e^{2\pi i  \tau_z/3}$, where $\bm{R}_i=\frac{4\pi}{3|q|^2}\bm{q}_i$, forming a triangular lattice.  
In the IVC crystal, the charge is modulated at the
Bragg wavevectors $\rho_{\bm{Q}_i}=\frac{1}{A}\sum_{\tau,\bm{k}}\langle c_{\tau,\bm{k}+\bm{Q}_i}^\dagger c_{\tau,\bm{k}}\rangle$, where $\bm{Q}_i=\bm{q}_{i+1}-\bm{q}_{i+2}$ 
(here, addition of the indices is taken modulo 3). 

Fig.~\ref{fig2}a and \ref{fig2}b show the ground state energy and the order parameters of IVC crystal and spiral phases as a function of carrier density at $\Delta_1=28$meV, along the black dashed line indicated in Fig. \ref{fig1}. 
We show the IVC order parameter (normalized per carrier),  $|\phi|=\tfrac{1}{|n|} \sqrt{\sum^3_{i=1} \phi^*_{\bm{q}_i} \phi_{\bm{q}_i}}$, which is non-zero in both IVC phases, and the CDW order parameter, $|\rho|=\tfrac{1}{|n|}\sqrt{\sum^3_{i=1} \rho^*_{\bm{Q}_i} \rho_{\bm{Q}_i}}$, which is nonzero only in the IVC crystal. Upon lowering the hole density, the system undergoes a transition from the half-metal phase where $\phi=\rho=0$ to the IVC crystal phase, where both order parameters are non-zero (Fig. \ref{fig2}b). Within our resolution, this transition seems to be continuous. 
The IVC crystal preserves $C_3$ symmetry ($|\phi_1|=|\phi_2|=|\phi_3|$).
At a lower hole density, the IVC crystal undergoes a first-order transition to an IVC spiral phase ($\phi\ne 0$, $\rho=0$). Upon further change of the doping towards charge neutrality, the system goes through a first order phase transition to a quarter metal with no IVC order.

Fig.~\ref{fig2}c and \ref{fig2}d show the spectral functions at the Fermi energy as a function of $\bm{k}$ in the IVC spiral and crystal phase. 
The non-interacting Fermi surfaces (FSs) of the two valleys (indicated by black lines) have an annular structure, with an outer hole and inner electron pockets. The wavevectors of the incommensurate IVC phases approximately nests the inner FS of one valley with the outer FS of the other valley, gapping parts of the FS. 

{\it Fluctuations and collective modes.--} 
Near the second order phase transition from the half metal phase to the IVC crystal, the intervalley susceptibility can be strongly enhanced.
We apply TDHF to extract the intervalley susceptibility 
when approaching the phase transition~\cite{SM}.
The dynamical intervalley susceptibility is given by 
\begin{equation}
    \chi_{\tau_{+},\bm{q},\omega}=-i\int_0^\infty dt \braket{[\hat{\phi}^\dagger(\bm{q},t), \hat{\phi}(\bm{q},0)]}e^{i\omega t},
    \label{eq:IVC_sus}
\end{equation} 
Fig. \ref{fig3}a shows the momentum space structure of the static susceptibility $\chi_{\tau_+,\omega = 0}$.
The IVC susceptibility is strongly enhanced along a contour connecting the inner and outer FS of the two valleys, with peaks located at a momentum $\bm{q}_0$ and its $C_3$ related momenta, where eventually the IVC instability develops.
Fig. \ref{fig3}b shows the enhancement of the static susceptibility $\chi_{\tau_+,\omega = 0}$ as a function of carrier density, approaching the transition to the IVC crystal. 
Interestingly, if the half-metal phase would remain the ground state at lower densities, it would develop strong susceptibilities even at small momenta; however, this is preempted by the transition to the IVC crystal phase.

Fig. \ref{fig3}c and \ref{fig3}d depict the real and imaginary parts of the dynamical susceptibility $\chi_{\tau_+,\bm{q}_0,\omega}$ vs. $\omega$. 
Upon approaching the phase transition, the peak in $\mathrm{Im}[\chi_{\tau_+,\bm{q}_0,\omega}]$ moves towards lower frequencies, signalling a softening of the IVC collective mode. Similarly, close to the transition, the zero-frequency peak in $\mathrm{Re}[\chi_{\tau_+,\bm{q}_0,\omega}]$ becomes increasingly stronger and narrower. 

In addition to the IVC mode, we compute the intravalley and intervalley magnon spectrum (See Fig.~\ref{fig10} in the SM~\cite{SM}). From the intravalley spectrum we extract the spin stiffness $\rho_s = 0.3$meV, which is surprisingly small in comparison to the bare interaction scale\footnote{In our model, neglecting intervalley Hund's coupling results in a $SU(2)\times SU(2)$ spin symmetry for independent rotations of each valley's spin. Thus, we find two decoupled Goldstone modes. Coupling the spins of the two valleys will result in a single Goldstone mode with roughly double the stiffness ($\rho_s\approx 0.6$meV), plus a gapped mode.}. 
Unlike the intravalley magnon, the intervalley magnon is not a Goldstone mode, and therefore it is gapped. However we find it has a relatively small gap of $0.35$meV, and a similar mass to that of the intravalley magnon. These findings may account for the large excess entropy seen recently in rhombohedral graphene at low $T$~\cite{holleis2024isospin}. 

{\it Superconductivity from intervalley fluctuations.--}
Based on our collective mode analysis in the vicinity of the transition from spin-polarized half-metal to IVC, we compute the leading superconducting instability, accounting for both the screened Coulomb interaction and the interaction induced by intervalley fluctuations. 
This analysis neglects particle-hole excitations in other channels, which give a subleading contribution close to the transition. 

The effective interaction becomes strongly frequency dependent near the transition, and the fermions near the FS become overdamped due to scattering off IVC fluctuations. For simplicity, we neglect both of these effects here, treating the full momentum structure of the effective interaction but neglecting its frequency dependence, and neglecting the fermion self-energy. Since the retardation and the fermion self-energy are expected to reduce $T_c$, our calculations pose an upper bound on the superconducting $T_c$ within our model.

\begin{figure}
    \centering
    \includegraphics[width=\linewidth]{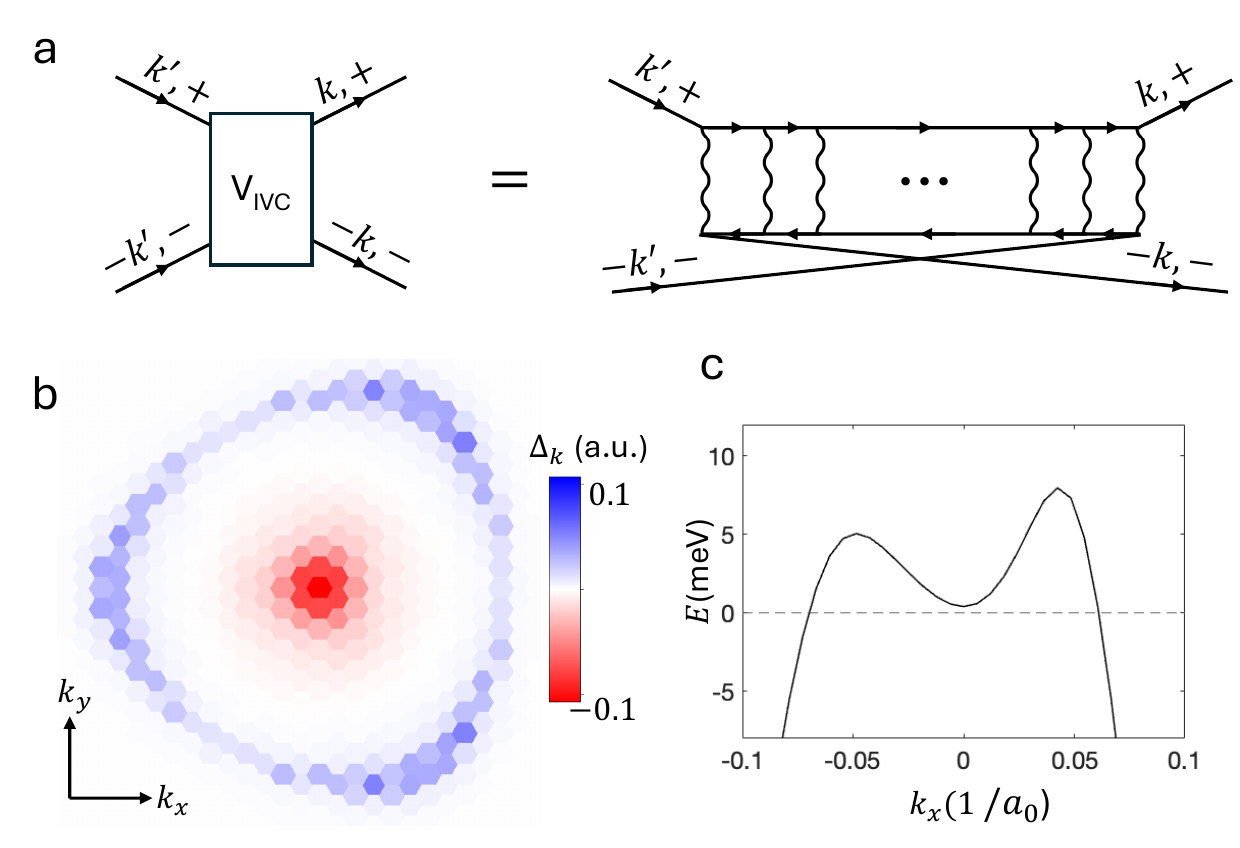}
    \caption{a) A diagrammatic representation of the IVC interaction. The IVC interaction is a sum over all such type of diagrams with two or more interaction lines.
    b) gap function for the largest eigenvalue of the linearized gap equation at displacement $\Delta_1 = 28$ meV and density $n=-1.03 \times {10}^{12} \text{cm}^{-2}$. 
    c) The Hartree-Fock spectrum at the corresponding displacement field and density. Although an annular FS is not formed, the states around $\bm{k}= 0$ are very close to the Fermi level, resulting in an $s_\pm $ order parameter.}
    \label{fig:superconductivity}
\end{figure}

The linearized gap equation is given by 
\begin{equation}    
    \Delta_{\bm{k}}=-\int \frac{d \bm{k}'}{(2\pi)^2} 
    \Delta_{\bm{k'}} 
    V_{\rm eff}(\bm{k},\bm{k'}) \frac{\tanh\left(\frac{\xi_{+,\bm{k'}}}{2 T}\right)}{2\xi_{+,\bm{k'}}},
    \label{eq:BCS}
\end{equation}
where $V_{\rm eff}(\bm{k},\bm{k}')=V_{+-}(\bm{k},-\bm{k};\bm{k'},-\bm{k'})+V_\text{IVC}(\bm{k},\bm{k'})$, 
$V_{+-}(\bm{k}_1,\bm{k}_2; \bm{k}'_1,\bm{k}'_2)$ being the Coulomb interaction matrix element between pairs of states  in the $+,-$ valleys, and $V_\text{IVC}$ is the interaction due to exchange of IVC fluctuations, illustrated diagrammatically in Fig. \ref{fig:superconductivity}a (see SM~\cite{SM} for details). $\xi_{+,\bm{k}}$ is the dispersion of holes in valley $+$.

We solve Eq. \eqref{eq:BCS} on a two-dimensional momentum grid.  
The resulting gap function for the largest eigenvalue is shown in Fig.~\ref{fig:superconductivity}b. 
Within Hartree-Fock, the FS in the half metal phase close to the transition consists of a single hole pocket in each valley, with a dispersion minimum very close to the Fermi energy at the center of the pocket (Fig.~\ref{fig:superconductivity}c). This dispersion minimum would become an additional pocket at lower hole density, giving rise to an annular FS, had the half metal not become unstable towards an IVC crystal\footnote{This feature of the calculation seems at odds with the experiment~\cite{zhou2021half}, where the instability of the half metal occurs in a regime where the FS is already annular. The presence of the inner pocket is expected to increase $T_c$}. Nevertheless, we find that $|\Delta_{\bm{k}}|$ is peaked at $\bm{k}=0$ where the incipient inner pocket resides. The gap has a sign-changing s-wave ($s_{\pm}$) structure within each valley. This is due to the structure of $V_{\rm IVC}$, which is repulsive in sign and peaked at a momentum transfer that connects the outer FS to the inner incipient pocket. At $n=-1.024\cdot10^{12}{\rm cm}^{-2}$ (about two percent away from the transition density to the IVC crystal), we find $T_c \approx 100$mK~\cite{SM}. 

{\it Discussion.--} In summary, we have studied the phase diagram near one of the superconducting regions in ABC graphene (SC2 of Ref. \cite{zhou2021superconductivity}), and found a transitions from a spin-polarized half metal to an IVC crystal, followed by a transition to an IVC spiral. The transition from the half metal to the IVC crystal is either of second order or very weakly first order, making it more difficult to detect in existing experiments that probe transport and compressibility. Nevertheless, the fluctuations associated with this transition are potentially important for superconductivity. We speculate that the first-order transition from the IVC crystal to the IVC spiral is the transition identified in Ref. \cite{zhou2021superconductivity} near SC2.   

The superconducting phase mediated by incommensurate IVC fluctuations is predicted to be of a sign-changing s-wave type, analogous to the one believed to occur in the iron-based superconductors~\cite{Mazin2008,Chubukov2008,Scalapino2012}. The basic mechanism is similar: IVC fluctuations induce a repulsive interaction that connects the inner and outer pockets of the annular FS, favoring a sign change in the order parameter. Within our calculations, superconductivity occurs in a regime where the inner pocket is very close to the Fermi energy but has not yet crossed it, analogously to the situation in certain iron based materials~\cite{Chen2015}.  

Our estimates of $T_c$ from incommensurate IVC fluctuations are high enough to account for the experimental findings in ABC graphene. 
We note, however, 
that the fermion self-energy and retardation effects, 
neglected in our calculation, may decrease
$T_c$. 
We leave a more complete theoretical treatment of this problem to future work. In addition, our calculation predicts that $T_c$ is appreciable in a narrow sliver of density around the symmetry-breaking transition.  
This seems consistent with the phase diagram in hBN-encapsulated ABC graphene~\cite{zhou2021superconductivity}. In AB graphene in proximity with WSe$_2$, however, superconductivity sometimes occurs in a broad range of density~\cite{zhang2023enhanced,holleis2024}, apparently at odds with the present mechanism. 

\begin{acknowledgments}
We thank I. Blinov, A. MacDonald, A. Young, and M. Zaletel for helpful discussions. 
This work was supported by NSF-BSF award DMR-2310312,  the European Research Council (ERC) under grant HQMAT (grant agreement No. 817799), CRC 183 of the Deutsche Forschungsgemeinschaft (subproject C02), and a research grant from
the Estate of Gerald Alexander.
T.H.\ acknowledges financial support by the 
European Research Council (ERC) under grant QuantumCUSP
(Grant Agreement No. 101077020). 
\end{acknowledgments}

\bibliography{ref}

\begin{widetext}
    
\section*{Supplemental Material for:
\texorpdfstring{\\}{ }
``Incommensurate inter-valley coherent states
\texorpdfstring{\\}{ }
in ABC graphene: collective modes and superconductivity''}

In this supplementary material, we give the details of the model and parameters used for ABC graphene, introduce the self-consistent Hatree-Fock approach, and derive the form of the collective modes within extended RPA. Finally, we comment on the solution of the linearied gap equation due to the enhancement of the IVC susceptibility.

\section{Model}
\label{Supp:Model}
The Hamiltonian used for our analysis is 
\begin{equation}
\mathcal{H}=\mathcal{H}_\text{free}+\mathcal{H}_{{\rm C}}=\sum_{\bm{k},\lambda}\psi_{\bm{k},\lambda}^{\dagger}{H}_{0}(\bm{k},\tau_z)\psi_{\bm{k},\lambda}+\mathcal{H}_{{\rm C}},
\label{eq:HFtotalHam}
\end{equation}
where $\lambda=(\tau_z,s_z)$ is a flavor index that includes both valley and spin, $H_0$ is the first-quantized non interacting Hamiltonian, $\mathcal{H}_\text{free}$ is the second-quantized non interacting Hamiltonian, and $\mathcal{H}_{{\rm C}}$ is the screened Coulomb interaction, taken to be valley-conserving. Throughout our calculation, we work in the approximation of a single active band, by projecting the Hamiltonian onto the third-lowest energy band of $\mathcal{H}_\text{free}$, appropriate for small hole doping away from charge neutrality.
\subsection{Non interacting Hamiltonian}
For the band structure of rhombohedral trilayer graphene, we use the
six band model of Ref.~\cite{zhang_band_2010}:
\begin{equation}
{H}_{0}(\bm{k},\tau_z)=\left(\begin{matrix}\Delta_{1}+\Delta_{2}+\delta & \frac{1}{2}\gamma_{2} & v_{0}\pi^{*} & v_{4}\pi^{*} & v_{3}\pi & 0\\
\frac{1}{2}\gamma_{2} & \Delta_{2}-\Delta_{1}+\delta & 0 & v_{3}\pi^{*} & v_{4}\pi & v_{0}\pi\\
v_{0}\pi & 0 & \Delta_{1}+\Delta_{2} & \gamma_{1} & v_{4}\pi^{*} & 0\\
v_{4}\pi & v_{3}\pi & \gamma_{1} & -2\Delta_{2} & v_{0}\pi^{*} & v_{4}\pi^{*}\\
v_{3}\pi^{*} & v_{4}\pi^{*} & v_{4}\pi & v_{0}\pi & -2\Delta_{2} & \gamma_{1}\\
0 & v_{0}\pi^{*} & 0 & v_{4}\pi & \gamma_{1} & \Delta_{2}-\Delta_{1}
\end{matrix}\right),\label{Ham6}
\end{equation}

where $\pi=\tau_z k_{x}+ik_{y}$ ($\tau_z=\pm$ corresponds to valleys $K$
and $K'$) and the Hamiltonian is written in the basis $(A_{1},B_{3},B_{1},A_{2},B_{2},A_{3})$,
where $A_{i}$ and $B_{i}$ label the two sublattices at layer $i$.
The velocities $v_{i}$ ($i=0,3,4$) are related to the microscopic
hopping parameters $\gamma_{i}$ by $v_{i}=\sqrt{3}a_{0}\gamma_{i}/2$,
where $a_{0}=2.46$\AA\  
is the lattice constant of monolayer graphene.
$\Delta_{1}$ is a potential difference between outer layers, which
is approximately proportional to the applied displacement field, while
$\Delta_{2}$ is the potential difference between the middle layer
and the average potential of the outer layers. Finally, $\delta$
is an on-site potential on $A_{1}$ and $B_{3}$. We have used the
following parameters: 
\begin{table}[h!]
    \centering
    \caption{Tight binding parameters}
    \label{tab:tight_binding}
    \begin{tabular}{|c|c|c|c|c|c|c|}
     $\gamma_{0}$& $\gamma_{1}$& $\gamma_{2}$& $\gamma_{3}$& $\gamma_{4}$& $\delta$& $\Delta_{2}$\\
     $3.1\,{\rm eV}$&  $0.38\,{\rm eV}$&  $-0.022\,{\rm eV}$& $-0.29\,{\rm eV}$& $-0.21\,{\rm eV}$& $-0.0105\,{\rm eV}$& $-0.0023\,{\rm eV}$
\end{tabular}\;.
\end{table}

Relative to the parameters in Ref.~\cite{zhou2021half}, we changed the values of  $\gamma_2$ and $\gamma_4$.
These parameters were adjusted to better fit the Hartree-Fock  phase diagram with the experimentally observed one. In particular, we were able to capture two features that are missed when using the parameters of Ref.~\cite{zhou2021half}: 1) The low density and low displacement field symmetry-unbroken phase with a twelve-fold degenerate FS~\cite{zhou2021half} (region 1 in Fig.~\ref{fig1}). 2) The sign of the slope of the boundary between the commensurate ($\bm{q}=0$) IVC and the valley polarized phases~\cite{arp2023intervalley} (regions 2,3 in Fig.~\ref{fig1}, respectively). 

\subsection{Screened Coulomb interaction}
The dual gate-screened interaction is given by $V_0(\bm{q})=2\pi e^2\tanh (q d)/(\epsilon q)$, where $d$ is the distance to the gates and $\epsilon$ is the dielectric constant. Inspired by the Thomas-Fermi approximation, we account in a phenomenological way for the effect of screening by the conduction electrons by modifying the interaction to
\begin{equation}
    V_{TF}(\bm{q})=\frac{V_0(\bm{q})}{1+q_{TF}(\epsilon/2\pi e^2)V_0(\bm{q})}.
\end{equation}
We take $\epsilon=8$ to mimic the dielectric constant of the encapsulating hexagonal Boron-Nitride and the remote ABC graphene bands, and $q_{TF} = 0.04\frac{1}{a_0}$ as mentioned in the main text. In comparison to previous works~\cite{arp2023intervalley,koh2024correlated}, where self-screening was accounted for through a phenomenological dielectric constant in the range of $\epsilon=20 \sim 40$ we find this method more physically motivated.

\begin{figure}
    \centering
    \includegraphics[width=0.6\linewidth]{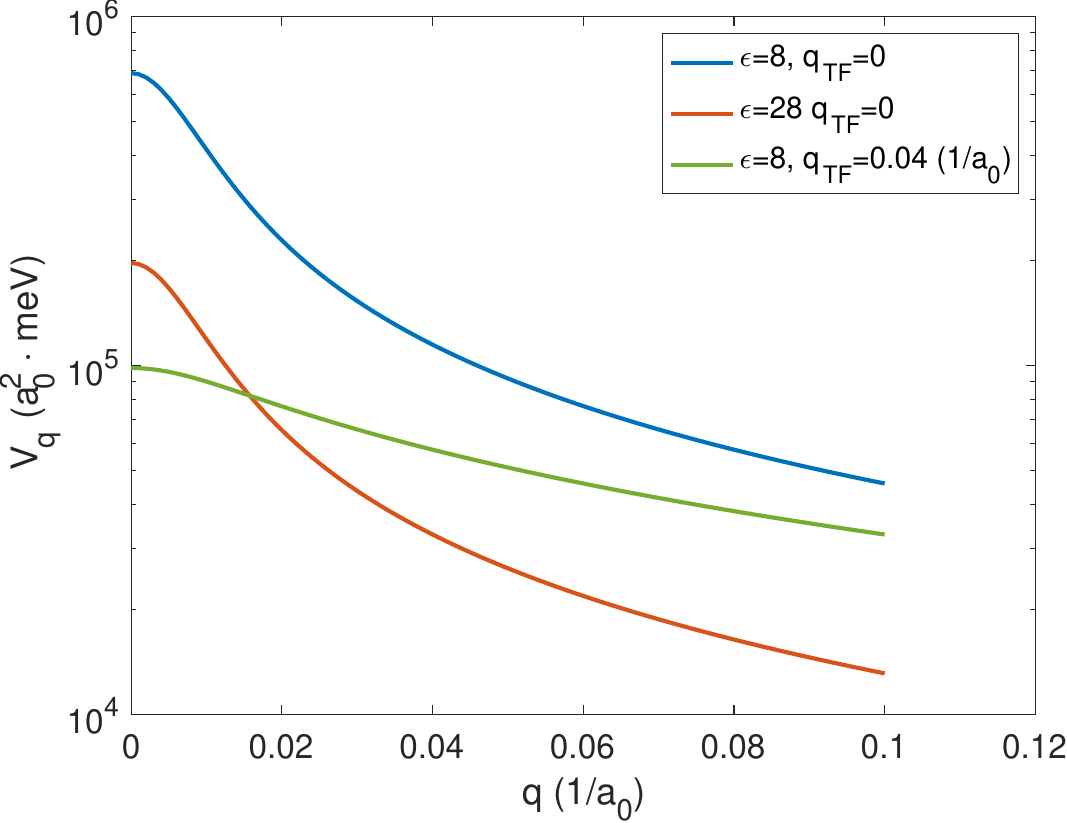}
    \caption{Interaction strength as function of momentum for interactions with and without self screening. In blue we show the case the bare Coulomb interaction with a dielectric constant that mimmics hBN effects. In red we show the interaction where self-screening is accounted for by a large dielectric constant (without any momentum dependence). In green we show the intearction where we account for self-screening using a phenomenological Thomas-Fermi vector. The red and green curves show a very distinct momentum dependence and the first way of approximating clearly underestimates large momentum interactions.}
    \label{fig:screened-interaction}
\end{figure}

The interaction part of the Hamiltonian is given by 
\begin{equation}
    \mathcal{H}_{{\rm C}}=\frac{1}{2A}\sum_{\lambda_1,\lambda_2,i,j}\sum_{\bm{k}_1,\bm{k}_2,\bm{k}_3,\bm{k}_4}V_{TF}(\bm{q}=\bm{k}_2-\bm{k}_3)\psi^\dagger_{i,\lambda_1,\bm{k}_1}\psi^\dagger_{j,\lambda_2,\bm{k}_2}\psi_{j,\lambda_2,\bm{k}_3}\psi_{i,\lambda_1,\bm{k}_4}\delta(\bm{k}_1+\bm{k}_2-\bm{k}_3-\bm{k}_4),
\end{equation}
where $\psi^\dagger_{i,\lambda,\bm{k}}$ creates an electron in layer/sublattice $i=1,\dots,6$, spin/valley flavor $\lambda$, and momentum $\bm{k}$. In order to project the interaction to a single active band, we use the transformation $\psi_{i,\lambda,\bm{k}}=\sum_n U_{i,n,\lambda}(\bm{k})\Tilde{\psi}_{n,\lambda,\bm{k}}$ from the sublattice basis to the band basis, and denote $a_{i,\lambda}(\bm{k})=U_{i,3,\lambda}(\bm{k})$ and $c_{\lambda,\bm{k}}=\Tilde{\psi}_{3,\lambda,\bm{k}}$ (here, $n=3$ is the partially filled band upon hole doping away from charge neutrality). The projector to the single partially filled band is given by 
\begin{equation}
    P=\prod_{\lambda,\bm{k}}\left[
    \prod_{n=1}^2\tilde{\psi}^\dagger_{n,\lambda,\bm{k}}\tilde{\psi}_{n,\lambda,\bm{k}}
    \prod_{n=4}^6\tilde{\psi}_{n,\lambda,\bm{k}}\tilde{\psi}^\dagger_{n,\lambda,\bm{k}},
    \right]
\end{equation}
and the projected interaction is 
\begin{multline}
    P\mathcal{H}_{{\rm C}}P=\\
    \frac{1}{2A}\sum_{\lambda_1,\lambda_2,i,j}\sum_{\bm{k}_1,\bm{k}_2,\bm{k}_3,\bm{k}_4}
    a_{i,\lambda_1}^*(\bm{k}_1)a_{j,\lambda_2}^*(\bm{k}_2)a_{j,\lambda_2}(\bm{k}_3)a_{i,\lambda_1}(\bm{k}_4)
    V_{TF}(\bm{q}=\bm{k}_2-\bm{k}_3)c^\dagger_{\lambda_1,\bm{k}_1}c^\dagger_{\lambda_2,\bm{k}_2}c_{\lambda_2,\bm{k}_3}c_{\lambda_1,\bm{k}_4}\delta(\bm{k}_1+\bm{k}_2-\bm{k}_3-\bm{k}_4)\\
    +\Delta\mu\sum_{\lambda_1,\bm{k}}c^\dagger_{\lambda_1,\bm{k}}c_{\lambda_1,\bm{k}},
\end{multline}
where $\Delta \mu$ is a shift in the chemical potential that we can absorb through a redefinition of the chemical potential.

\subsection{Gauge Fixing}
It is worth noting that while the single-band projector and the projected interaction are gauge-invariant operators, the band creation and annihilation operators ($c^\dagger_{\lambda,\bm{k}}, c_{\lambda,\bm{k}}$) by themselves are not gauge-invariant. This will become important later when we define the order parameter and the intervalley susceptibility in terms of these operators. We fix the gauge by demanding that the overlap of the band annihilation operator $c_{\lambda,\bm{k}}$ with the annihilation of an electron in the $B3$ sublattice $\psi_{6,\lambda,\bm{k}}$ is real and positive (i.e., $U_{6,3,\lambda}$ is real and positive). This choice is motivated by the fact that under large displacement field the valence band creation and annihilation operators have large overlap with the creation and annihilation operators in the $B3$ sublattice.

\section{Self-consistent Hartree-Fock calculations}
\subsection{Slater Determinant {\it Ansatz}}
For the self-consistent Hartree-Fock calculations, we use an {\it ansatz} that allows for a non-zero expectation value of the IVC order parameter $\phi_{\bm{q}}$ at a given momentum $\bm{q}_1$ and its $C_3$ related momenta ($\bm{q}_3=R_{2\pi/3}\cdot \bm{q}_2=R_{4\pi/3}\cdot \bm{q}_1$).
As discussed in the main text, this set of momenta naturally defines a set of three CDW ordering vectors $\bm{Q}_i=\bm{q}_{i+1}-\bm{q}_{i+2}$ at which our ansatz allows for a finite CDW order parameter $\rho_{\bm{Q}_i}$. Additionally, we allow for CDW order at higher harmonics of these momenta, up to a cutoff momentum $\Lambda$. We denote the set of momenta in the reciprocal lattice included in our calculation by $A_{\text{CDW}}$, defined as $A_{\text{CDW}}=\{\bm{Q}=n_1\bm{Q}_1+n_2\bm{Q}_2+n_3\bm{Q}_3 \; | \;  n_i\in\mathbb{Z},|\bm{Q}|<\Lambda \}$, and the set of IVC ordering vectors at momenta which are equal to $\bm{q}_1$ plus a CDW ordering vector $A_\text{IVC}=\{\bm{q}=\bm{q}_1 +\bm{Q} \; | \;  \bm{Q}\in A_\text{CDW} \}$. Such an ansatz is automatically realized by defining a mini Brillouin zone (mBz) spanned by $\bm{Q}_1$ and $\bm{Q}_2$, and allowing for hybridization between all single particle states with the same lattice-momentum. In order to capture the correct IVC momenta, we need to choose the mBz centers for the two valleys to be shifted by $\bm{q}_1$ (which is equal to the $K$ vector of the mBz by construction) as depicted in Fig.~\ref{fig:CDW_BZ}. 

The resulting variational Hamiltonian is 
\begin{equation}
    \mathcal{H}_\text{var}=\mathcal{H}_\text{free}+\sum_{\lambda_1,\lambda_2}\sum_{\bm{G}_1,\bm{G}_2}\sum_{\bm{k}\in \text{mBz}}t_{\lambda_1 \lambda_2}^{\left( \bm{G}_1, \bm{G}_2\right)} \left( \bm{k} \right) c^\dagger_{\lambda_1,\bm{k}_1}c_{\lambda_2, \bm{k}_2},
\end{equation}
 where $\lambda=(\tau_z,s_z)$ is a flavor index that includes both valley and spin, $\bm{G}_1,\bm{G}_2\in A_\text{CDW}$ are mBZ primitive vectors, $\bm{k}$ is lattice momentum (and its summation is in the first mBZ), $\bm{k}_1=\bm{k}+\bm{G}_1+\tau_z^{(1)}\bm{q}_1/2$ and $\bm{k}_2=\bm{k}+\bm{G}_2+\tau_z^{(2)}\bm{q}_1/2$.

Defining $u_{\lambda_1\lambda_2;ij}^{\left( \bm{G}_1, \bm{G}_2\right)}(\bm{k})=a^*_{i,\lambda_1}(\bm{k}_1)a_{j,\lambda_2}(\bm{k}_2)$ and $\Phi_{\lambda_1\lambda_2}^{\left( \bm{G}_1, \bm{G}_2\right)}(\bm{k})=\left\langle c^\dagger_{\lambda_1,\bm{k}_1}c_{\lambda_2,\bm{k}_2}\right\rangle_\text{var}$ where $\bm{k}_1,\bm{k}_2$ are defined as in the previous paragraph, the mean-field equation is given by
\begin{multline}
    t_{\lambda_1 \lambda_2}^{\left( \bm{G}_1, \bm{G}_2\right)}(\bm{k})=\delta_{\lambda_1 \lambda_2}\frac{V_{TF}(\delta\bm{G})}{A}\sum_i u_{\lambda_1 \lambda_1;ii}^{\left( \bm{G}_1, \bm{G}_2\right)}(\bm{k})\left[ \sum_{\bm{G}_3}\sum_{\bm{k}'}\sum_{\lambda_3}\sum_{j} u_{\lambda_3\lambda_3;ii}^{\left( \bm{G}_3, \bm{G}_3+\delta\bm{G}\right)}(\bm{k}')\Phi_{\lambda_3\lambda_3}^{\left( \bm{G}_3, \bm{G}_3+\delta\bm{G}\right)}(\bm{k}') \right]\\
    -\sum_{ij}u_{\lambda_1 \lambda_2;ji}^{\left( \bm{G}_1, \bm{G}_2\right)}(\bm{k})\sum_{\bm{G}_3}\sum_{\bm{k}'}
    \frac{V_{TF}(\bm{k}+\bm{G}_2-\bm{k}'-\bm{G}_3)}{A} u_{\lambda_2\lambda_1 ;ij}^{\left( \bm{G}_3, \bm{G}_3+\delta\bm{G}\right)}(\bm{k}') \Phi_{\lambda_2\lambda_1}^{\left( \bm{G}_3, \bm{G}_3+\delta\bm{G}\right)}(\bm{k}'),
\end{multline}
where $\delta\bm{G}=\bm{G}_1-\bm{G}_2$.

\subsection{Solution of the Self Consistent Equation}

We solve the self-consistent equations using the optimal damping approach~\cite{defranceschi2012mathematical}. In the following we give a brief description of the algorithm. 

The variational Slater-determinant state is uniquely determined by the set of correlators $\{\Phi_{\lambda_1\lambda_2}^{\left( \bm{G}_1, \bm{G}_2\right)}(\bm{k})\}$ which we denote for simplicity by $\hat{\Phi}$. At each iteration $\tau$ we solve the variational Hamiltonian defined through the current state $\mathcal{H}_\text{var}(\hat{\Phi}_\tau)$ and calculate the correlators in its ground state, which we denote by $\hat{\Phi}'_\tau$. We then set $\hat{\Phi}_{\tau+1}$ to be used in the next iteration according to
\begin{equation}
    \hat{\Phi}_{\tau+1}=\hat{\Phi}(r)=(1-r)\hat{\Phi}_\tau +r\hat{\Phi}'_\tau.
    \label{eq:Phi}
\end{equation} 
Eq. \eqref{eq:Phi} is a convex sum of $\hat{\Phi}_{\tau}$
and the ground state of $\mathcal{H}_\text{var}$, where $r\in[0,1]$ is the update rate. It can be shown that the energy $\braket{\mathcal{H}}_{\hat{\Phi}(r)}$ is a second order polynomial of $r$, where the coefficients of the polynomial can be calculated efficiently. Therefore, at each time-step we find the polynomial and choose $r=\arg \min \bigl[ \braket{\mathcal{H}}_{\hat{\Phi}(r)} \bigr]$. This method has two advantages: 1) It converges much faster than methods with a constant update rate, 2) It can be proven to converge to a local minima of $E_\text{var}=\braket{\mathcal{H}}_{\hat{\Phi}}$ (see \cite{defranceschi2012mathematical} for a proof).

We solve this self-consistent equation for many different initial conditions with different broken symmetries, and compare their energy directly, in order to insure convergence to the best variational state with high probability. In proximity to phase transitions we take special care by using transfer-learning, i.e., we use converged states from across the transition as trial states to avoid convergence towards the wrong phase due to a bad starting point.

\begin{figure}
    \centering
    \includegraphics[width=0.4\linewidth]{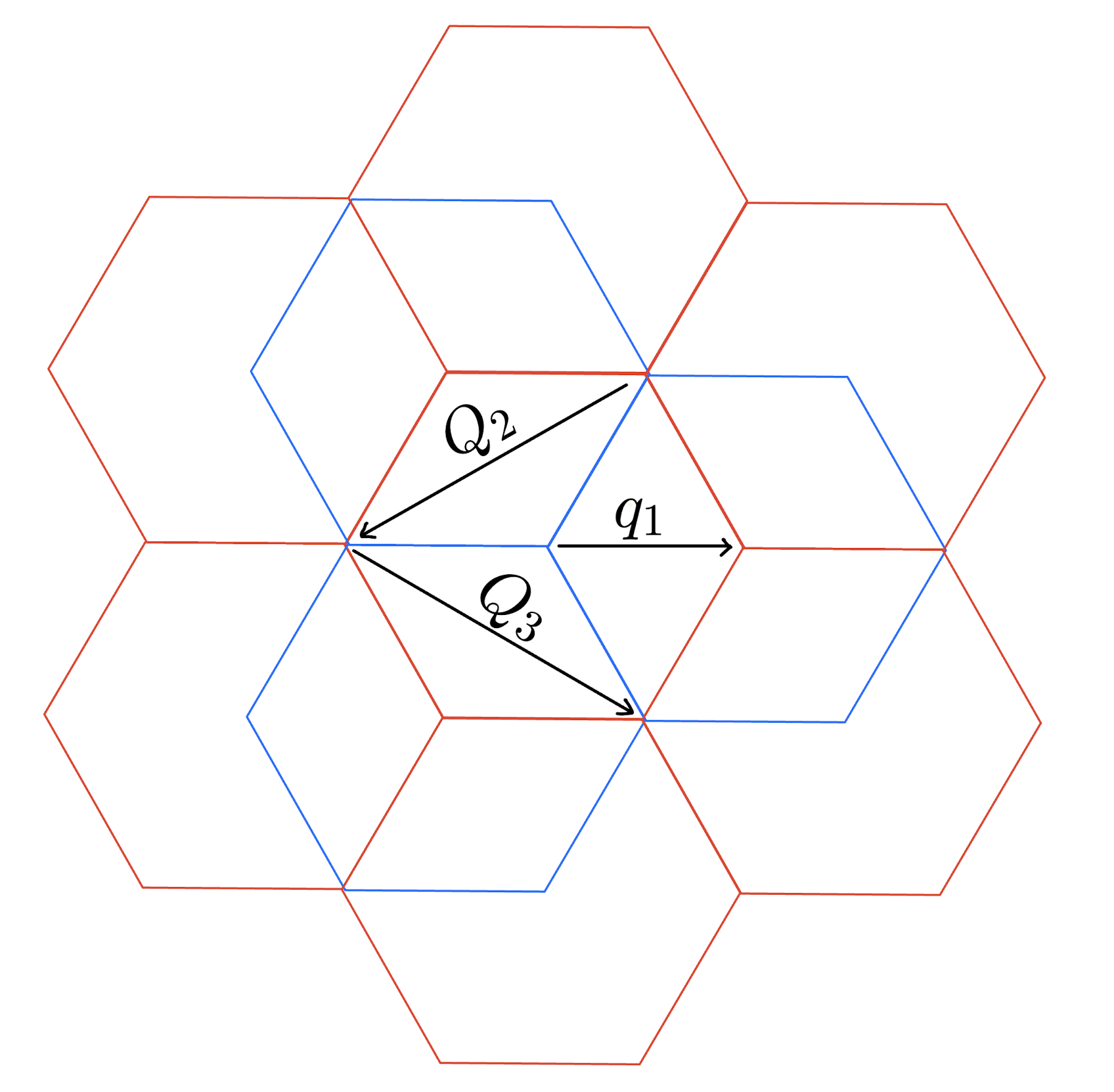}
    \caption{The mini Brillouin zone choice for the $\tau=-$ valley (red) and the $\tau=+$ valley (blue). The relative mBZ center shift between the valleys corresponding to the allowed intervalley coherent ordering vector, and the primitive vectors of the mBZ corresponding to the allowed charge density ordering wave-vector are depicted in black.}
    \label{fig:CDW_BZ}
\end{figure}

\section{Collective modes}
\subsection{Generalized RPA (Time dependent Hartree Fock)}
\begin{figure}
    \centering
    \includegraphics[width=0.8\linewidth]{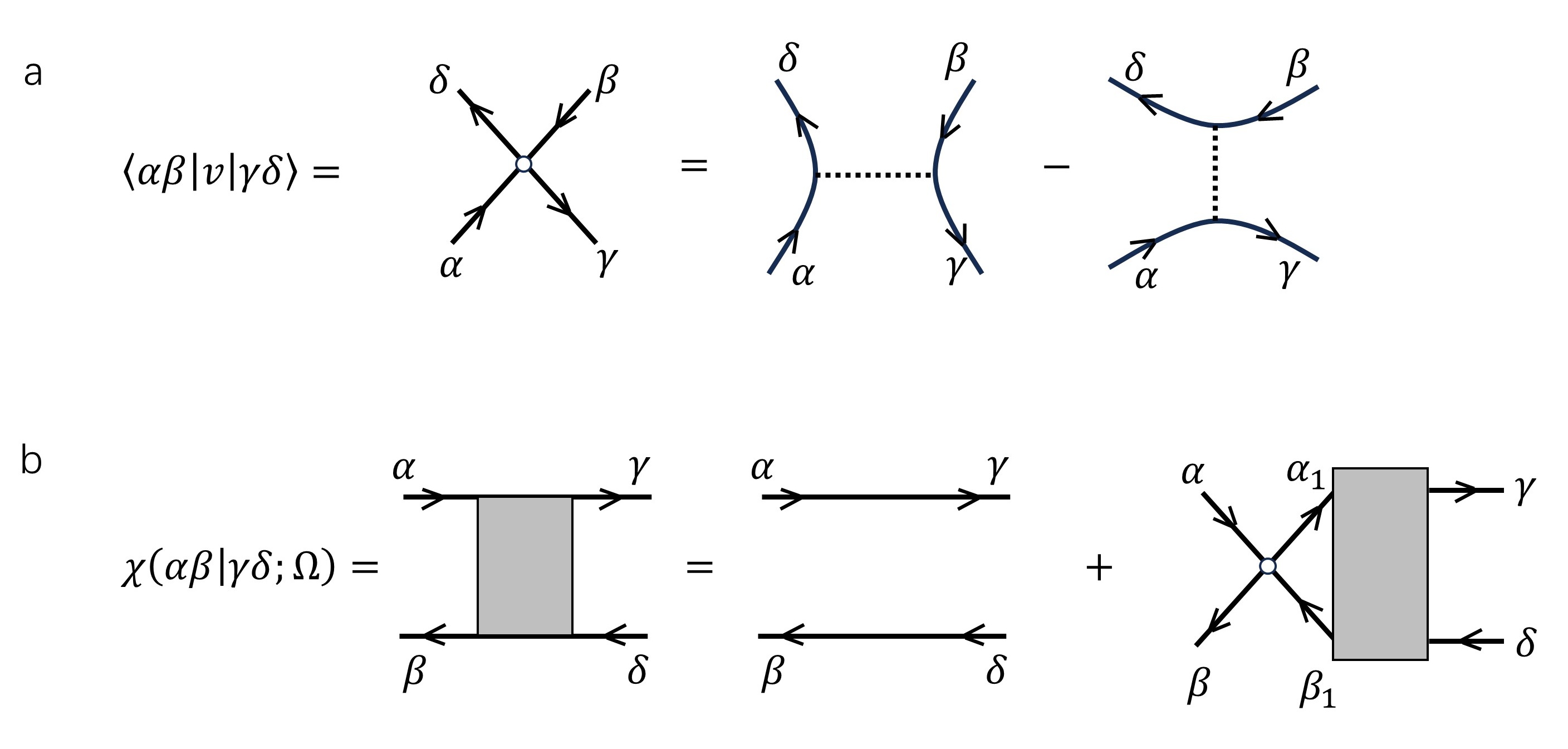}
    \caption{Diagrams. a. Coulomb interaction vertex, where $\alpha,\beta,\gamma,\delta$ labels both flavor index and momentum index. The Coulomb vertex gives both direct and exchange terms. b. The generalized RPA self-consistent equation. }
    \label{fig6}
\end{figure}
We are interested in the particle-hole correlation function (susceptibility) given the Hartree Fock ground states.
The susceptibility is defined as:
\begin{equation}
    \chi(\alpha\beta|\gamma\delta;\Omega) = \braket{(c^\dagger_{\alpha} c_\beta)_\Omega (c^\dagger_{\delta} c_\gamma)_{-\Omega} },
    \label{eq:chi}
\end{equation}
where $\alpha,\beta,\gamma,\delta$ label the Hartree-Fock orbitals. In the half metal phase, these include the momentum and flavor index: $\alpha = (\lambda_\alpha,\mathbf{k}_\alpha)$. 
Assuming the interaction has the form:
\begin{equation}
   H_{C} = \sum_{\alpha\beta\gamma\delta} \braket{\alpha\beta|v|\gamma\delta} c^{\dagger}_{\alpha}c^{\dagger}_{\beta}c_{\gamma}c_{\delta},
\end{equation}
where we include both the form factors and screened Coulomb interactions into the matrix element $\braket{\alpha\beta|v|\gamma\delta}$:
\begin{equation}
    \braket{\alpha\beta|v|\gamma\delta} = \sum_{ij} \Big( a^{*}_{i,\alpha} a^{*}_{j,\beta} a_{j,\gamma} a_{i,\delta} V_{\bm{k}_\alpha-\bm{k}_\delta}\delta_{\lambda_\alpha\lambda_\delta}\delta_{\lambda_\beta\lambda_\gamma} - 
    a^{*}_{i,\alpha} a^{*}_{j,\beta} a_{j,\gamma} a_{i,\delta} V_{\bm{k}_\alpha-\bm{k}_\gamma}\delta_{\lambda_\alpha\lambda_\gamma}\delta_{\lambda_\beta\lambda_\delta} \Big)\delta_{\mathbf{k}_{\alpha}+\mathbf{k}_{\beta},\mathbf{k}_{\gamma}+\mathbf{k}_{\delta}}
    \label{eq:coulomb_matrix_element}
\end{equation}
This interaction vertex is shown in Fig. \ref{fig6}a, giving both direct and exchange terms. Note that $\braket{\alpha\beta|v|\gamma\delta}= -\braket{\alpha\beta|v|\delta\gamma} = -\braket{\beta\alpha|v|\gamma\delta}$. 
Within the random phase approximation (RPA), $\chi$ satisfies the self-consistent equation shown diagrammatically in Fig. \ref{fig6}b, written as:
\begin{equation}\label{eq11}
   \chi(\alpha\beta|\gamma\delta;\Omega) = \frac{n(\varepsilon_\alpha)-n(\varepsilon_\beta)}{i\Omega + \varepsilon_\beta - \varepsilon_\alpha}\Big(\delta_{\alpha\gamma}\delta_{\beta\delta} + \sum_{\alpha_1\beta_1}\braket{\alpha\beta_1|v|\beta\alpha_1} \chi(\alpha_1\beta_1|\gamma\delta;\Omega)\Big),
\end{equation}
where $\Omega$ is the Matubara frequency and $n(\varepsilon)$ is the Fermi function. 
Specializing to $T=0$, we can further simplify the above equation by noticing that  $\chi(\alpha\beta|\gamma\delta;\Omega)= 0$ when $n(\varepsilon_\alpha)-n(\varepsilon_\beta)= 0$, i.e., if both states are occupied or both states are empty.
Following Ref. \cite{negele2018quantum}, we label the occupied states as $A,B$ and empty states as $a,b$. 
We can divide eq. \ref{eq11} into two parts when $\alpha=a,\beta=A$ and $\alpha=A,\beta=a$:
\begin{equation}
   \chi(aA|\gamma\delta;\Omega) = \frac{-1}{i\Omega + \varepsilon_A - \varepsilon_a}
   \Big(\delta_{a\gamma}\delta_{A\delta} +\sum_{bB}\braket{aB|v|Ab} \chi(bB|\gamma\delta;\Omega)
   +\sum_{Bb}\braket{ab|v|AB} \chi(Bb|\gamma\delta;\Omega)
   \Big)
\end{equation}
\begin{equation}
   \chi(Aa|\gamma\delta;\Omega) = \frac{-1}{i\Omega + \varepsilon_a - \varepsilon_A}
   \Big(\delta_{A\gamma}\delta_{a\delta} + \sum_{Bb}\braket{Ab|v|aB} \chi(Bb|\gamma\delta;\Omega) 
   + \sum_{bB}\braket{AB|v|ab} \chi(bB|\gamma\delta;\Omega)
   \Big)
\end{equation}
We can rewrite the above two equations as
\begin{equation}
\sum_{bB}
\left(
\begin{array}{cc}
(i\Omega + \varepsilon_A-\varepsilon_a)\delta_{AB}\delta_{ab} + \braket{aB|v|Ab} & \braket{ab|v|AB} \\
\braket{BA|v|ba} & (-i\Omega + \varepsilon_A-\varepsilon_a)\delta_{AB}\delta_{ab} + \braket{bA|v|Ba}   \\
\end{array}
\right)  
\left(
\begin{array}{c}
\chi(bB|\gamma\delta) \\
\chi(Bb|\gamma\delta)  \\
\end{array}
\right) = 
\left(
\begin{array}{c}
-\delta_{a\gamma} \\
-\delta_{A\delta}  \\
\end{array}
\right).
\end{equation}
In matrix form, this equation can be written as:
\begin{equation}\label{eq16}
    (i\Omega\Sigma_z - H)\chi = 1,
\end{equation}
where $H$ and $\Sigma_z$ are given by
\begin{equation}
    H = \left(
\begin{array}{cc}
(\varepsilon_a-\varepsilon_A)\delta_{AB}\delta_{ab} -\braket{aB|v|Ab} & -\braket{ab|v|AB} \\
-\braket{BA|v|ba} & (\varepsilon_a-\varepsilon_A)\delta_{AB}\delta_{ab} - \braket{bA|v|Ba}   \\
\end{array}
\right),  
\end{equation}
\begin{equation}
    \Sigma_z = \left(
\begin{array}{cc}
1 & 0 \\
0 & -1 \\
\end{array}
\right).  
\end{equation}
The diagonal block is composed of $(\varepsilon_a-\varepsilon_A)\delta_{AB}\delta_{ab}$ and $\braket{aB|v|bA}$. The first part describes particle-hole excitations based on Hartree-Fock ground states, and the second part is the Coulomb coupling between particle-hole excitations.
The off-diagonal block $\braket{AB|v|ba}$ is the coupling between particle-hole excitations and de-excitations, and induces a mixing of particle-hole excitations of the HF Hamiltonian in the actual ground-state.
For example, for the intervalley excitations in the half metal phase, the occupied states $A,B$ is denoted by $A=(\bm{k},K),B=(\bm{k'},K)$ and the empty state $a=(\bm{k+q},K'),b=(\bm{k'+q},K')$, we can write the coupling between particle-hole excitations as:
\begin{equation}
\begin{aligned}
    \braket{aB|v|bA} &= \braket{a=\bm{k+q},K';B=\bm{k'},K|v|b=\bm{k'+q},K';A=\bm{k},K} \\
    &= \sum_{ij} n_{\bm{k},K} n_{\bm{k'},K}(1-n_{\bm{k+q},K'})(1-n_{\bm{k'+q},K'}) V_{\bm{k-k'}} a^*_{i,K,\bm{k'}}a_{i,K,\bm{k}} a^*_{j,K',\bm{k+q}}a_{j,K',\bm{k'+q}},
    \end{aligned}
\end{equation}
where $\bm{q}$ is the momentum of collective modes.
For the off-diagonal block, we can write the coupling between intervalley excitation and de-excitation as:
\begin{equation}
\begin{aligned}
    \braket{ab|v|AB} &= \braket{a=\bm{k+q},K';B=\bm{k'},K'|v|b=\bm{k'-q},K;A=\bm{k},K} \\
    &= \sum_{ij} n_{\bm{k},K} n_{\bm{k'},K'}(1-n_{\bm{k+q},K'})(1-n_{\bm{k'-q},K}) V_{\bm{k+q-k'}} a^*_{i,K',\bm{k+q}}a_{i,K,\bm{k'}} a^*_{j,K,\bm{k'-q}}a_{j,K,\bm{k}}.
    \end{aligned}
\end{equation}
To obtain the collective mode spectrum and the dynamical susceptibilities, we multiply Eq. \eqref{eq16} by $\Sigma_z$ and diagonalize the matrix $\Sigma_zH$ as $\Sigma_zH = \hat{O}\Sigma_z\hat{w}\hat{O}^{-1}$. $\hat{\omega}$ is a diagonal matrix that contains the collective modes spectrum. Note since $\Sigma_z H$ is not hermitian, imaginary $\omega$'s are possible, the appearance of which indicate an instability of the Hartree-Fock state. For example, as shown in Fig. \ref{fig8}, when crossing the phase transition to the IVC crystal phase (white line in Fig. \ref{fig8}a), the half metal develops an instability, signaled by a non-zero imaginary part of the collective mode frequencies.
In terms of $\hat{O}$ and $\hat{\omega}$, the susceptibility matrix $\chi(\Omega)$ can be expressed as:
\begin{equation}
    \chi(\Omega) = -\hat{O} (i\Omega-\Sigma_z \hat{\omega})^{-1} O^{-1}\Sigma_z.
\end{equation}
\begin{figure}
    \centering
    \includegraphics[width=0.6\linewidth]{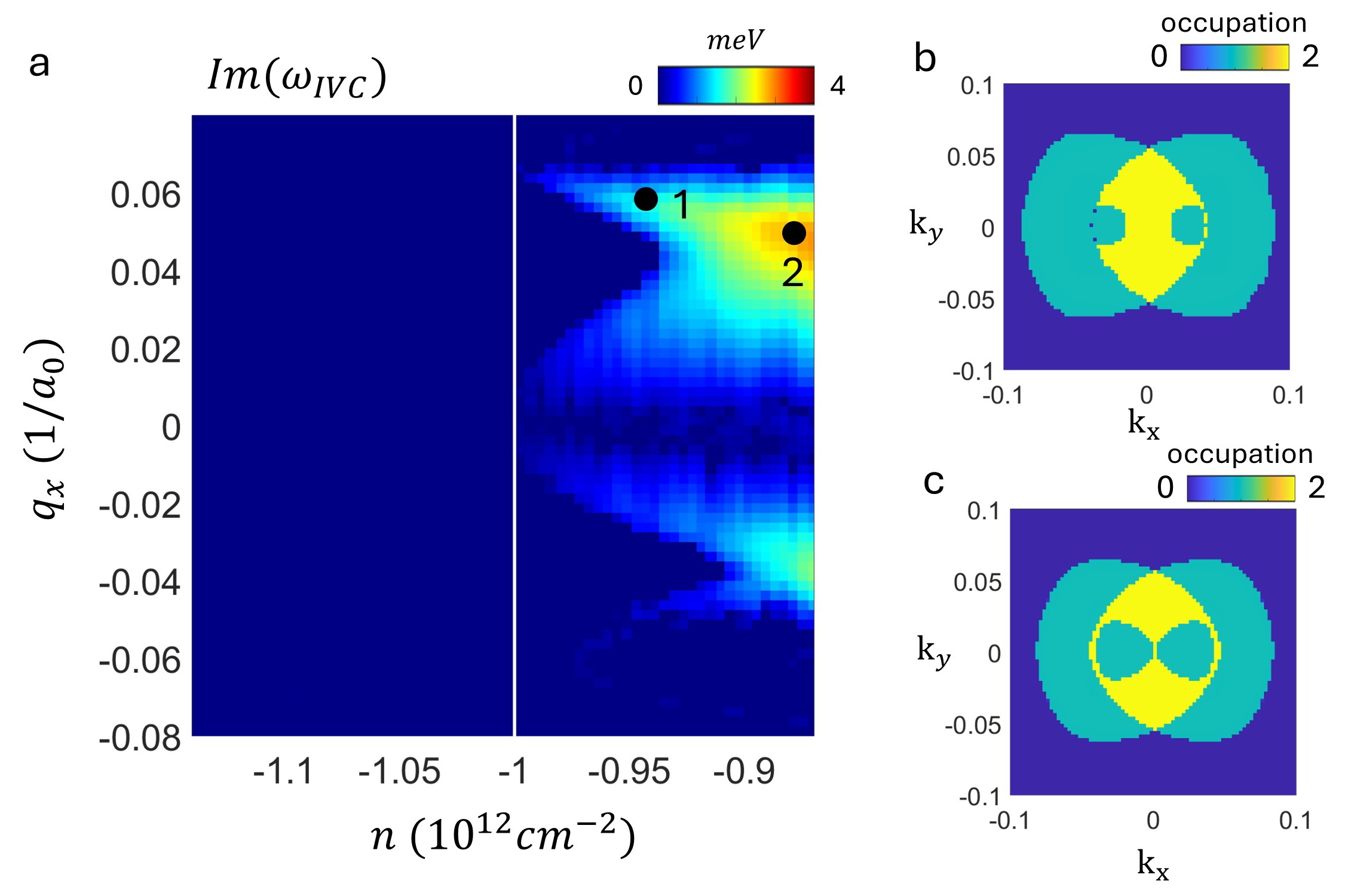}
    \caption{a. Imaginary part of the interlayer valley collective mode frequencies, ${\rm Im}(\omega_{IVC})$, calculated in the half metallic state. The white line marks the onset of a non-zero imaginary part, which signals an instability towards an incommensurate IVC phase. b, c. Hole occupations {of the half-metalic phase} at point 1 and 2 in panel a, with the corresponding shift vector $q_x$ that maximize the ${\rm Im}(\omega_{IVC})$ at a given density.}
    \label{fig8}
\end{figure}

\subsection{IVC Susceptibility}
The IVC susceptibility defined in Eq.~\ref{eq:IVC_sus} can be expressed using the susceptibility matrix (Eq.~\ref{eq:chi}).
We start by defining the Matsubara counterpart of the IVC susceptibility
\begin{align}
    \tilde{\chi}_{\tau_+,\bm{q},\Omega} = \braket{\hat{\phi}_{\bm{q},\Omega} \hat{\phi}_{\bm{q},\Omega}^\dagger} &= \frac{1}{A^2}\sum_{\bm{k},\bm{k}'}\braket{(c^\dagger_{+,\bm{k}+\bm{q}} c_{-,\bm{k}})_\Omega (c^\dagger_{-,\bm{k}'-\bm{q}} c_{+,\bm{k}'})_{-\Omega} }\notag\\
    &=\frac{1}{A^2}\sum_{\bm{k},\bm{k}'}\chi\left[(+,\bm{k}+\bm{q}),(-,\bm{k})|(+,\bm{k}'),(-,\bm{k}'-\bm{q});\Omega\right],
\end{align}
where we used the notation $(\pm,\bm{k})$ for a quasi particle in valley $\pm$ with momentum $\bm{k}$.
Through analytical continuation we can find the retarded susceptibility in Eq.~\ref{eq:IVC_sus} as
\begin{equation}
    \chi_{\tau_+,\bm{q},\omega}=\tilde{\chi}_{\tau_+,\bm{q},\Omega=-i\omega+\eta}.
\end{equation}

\subsection{Other collective modes}
Here we present other collective modes in the system: the plasmon, intra-valley magnon (involving a spin flip within a valley), and inter-valley magnon (involving a transfer of an electron from valley $+$ to valley $-$ with a flipped spin), all in the half metallic spin-polarized phase (Fig.~\ref{fig10}).
Figs.~\ref{fig10}a, b show the static charge susceptibility and corresponding line cuts as a function of $q_x$ at displacement $\Delta_1=28$meV and the same densities range as in Fig.~\ref{fig3}.
The charge susceptibility $\chi_C$ shows an increase upon approaching the transition. However, it remains non-singular, which suggests that the transition to the IVC crystal is not driven by a tendency towards charge ordering. 
Additionally, we show the spectrum for both types of magnons, at carrier density of $n=1.2\times 10^{12}{\rm cm}^{-2}$ and displacement $\Delta_1=28$meV, where the half-metal is stable. 
The intra-valley magnon (Fig. \ref{fig10}c, d) is gapless with a quadratic dispersion, as required by Goldstone's theorem; the inter-valley magnon (Fig. \ref{fig10}e, f) is gapped.

Fig. \ref{fig11} shows the dynamic charge susceptibility as a function of $\omega$ and carrier density at displacement $\Delta_1=28$meV and the same density range as in Fig.~\ref{fig3}. 
Here, we plot $\chi_C$ at the same momentum $q=q_0$ as in Fig. \ref{fig3} in the main text.
Overall, we find a gradual increase of the dynamic susceptibility around zero frequency. 

\begin{figure}
    \centering
    \includegraphics[width=0.9\linewidth]{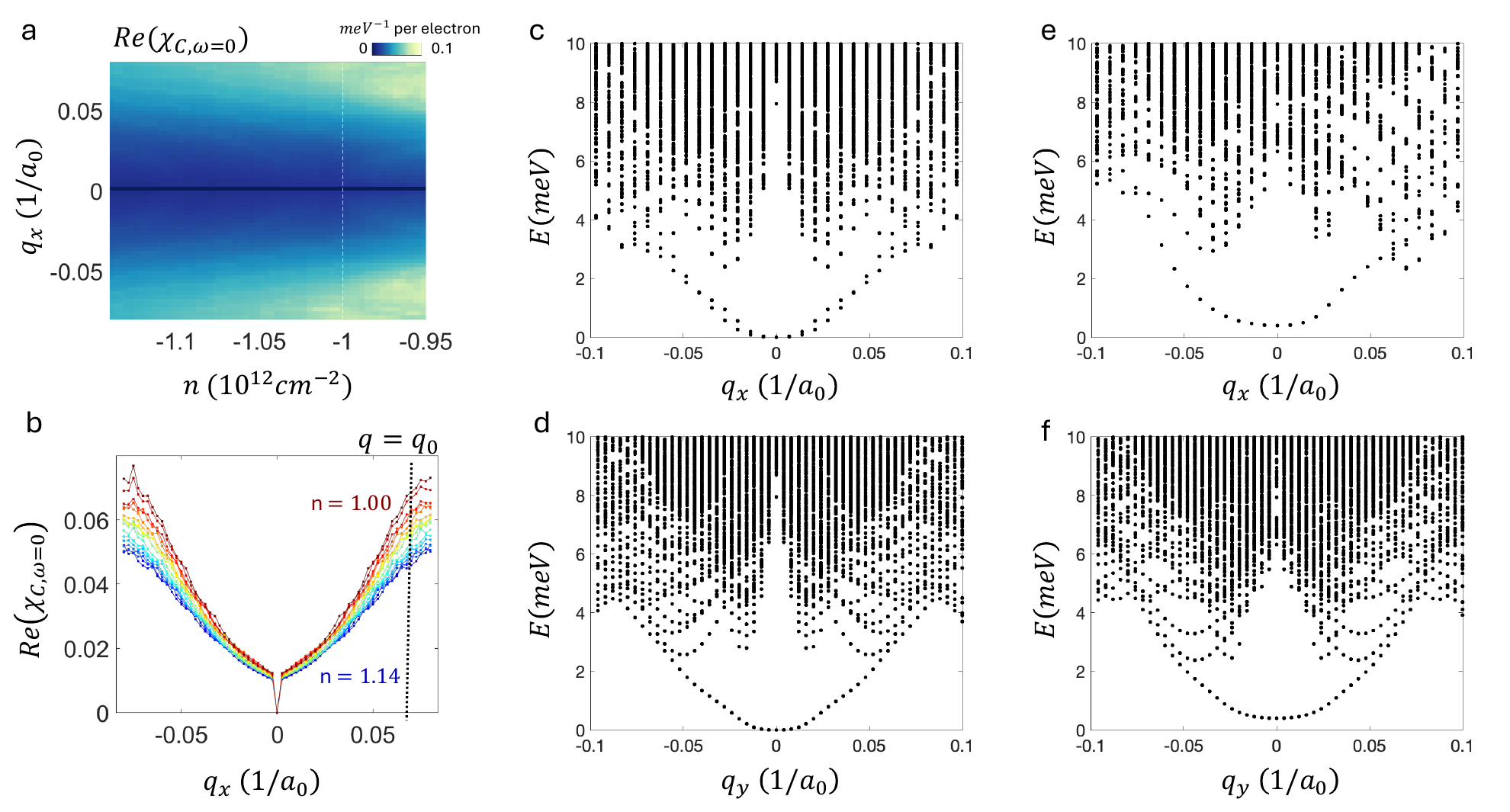}
    \caption{a) Charge susceptibility $\chi_{C}$ in the half-metal phase as a function of $q_x$ and carrier density at displacement $\Delta_1=28$meV and the same density range as in Fig.~\ref{fig3}.
    The dashed line marks the phase boundary between the half metal and the IVC crystal. b) Line cuts of $\chi_{C}$ at $\omega = 0$ as a function $q_x$. We mark the same $q_0$ as in Fig.\ref{fig3}b.  c,d) Intravalley magnon spectrum along 1d momentum line cuts with $q_y=0$ and $q_x=0$, respectively. e,f) Same as c,d) for intervalley magnon.}
    \label{fig10}
\end{figure}

\begin{figure}
    \centering
    \includegraphics[width=0.6\linewidth]{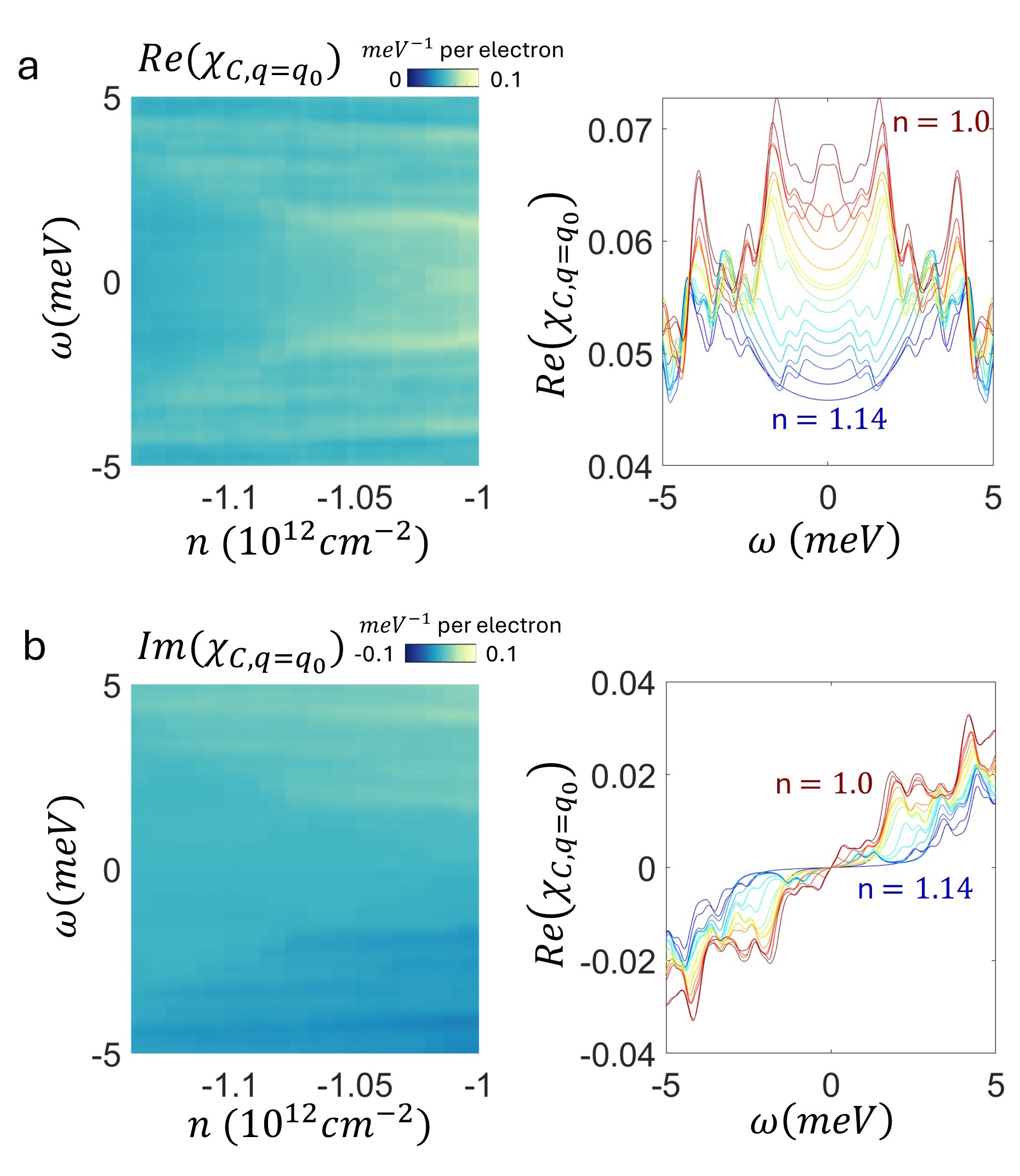}
    \caption{a) and b) The real and imaginary part of charge susceptibility $\chi_{C}$ at momentum $q = q_0$ as a function of $\omega$ and carrier density. Displacement field and carrier density range are the same as in Fig.~\ref{fig3}. Right panel plots the corresponding line cuts as a function of $\omega$.}
    \label{fig11}
\end{figure}

\subsection{Spin Stiffness}

To compute the spin stiffness of the half metal phase, we use the fact that this phase is fully spin-polarized, such that the total spin density is $|S|=\frac{|n|}{2}$. The Lagrangian density that describes the spin dynamics is written as
\begin{equation}
    \mathcal{L}=\bm{\mathcal{A}}(\hat{s})\cdot \Dot{\hat{s}}-\frac{\rho_s}{2}(\nabla \hat{s})^2 \approx \frac{|n|}{4}(\hat{s}_x \Dot{\hat{s}}_y - \hat{s}_y \Dot{\hat{s}}_x) -\frac{\rho_s}{2}\left( (\nabla \hat{s}_x)^2 + (\nabla \hat{s}_y)^2 \right),
\end{equation}
where $\hat{s}$ is the spin orientation vector of unit length, $\bm{\mathcal{A}}(\theta,\varphi) = \frac{|n|}{2} \tan \left( \frac{\theta}{2} \right) \hat{\varphi}$ is the Berry connection and $|n|$ is the carrier density. The approximation follows from an expansion around $\hat{s}=\hat{z}$. The spin-wave dispersion is thus $\omega_{\bm{q}}=\frac{2\rho_s}{|n|} q^2$. By fitting the dispersion of the intra-valley magnon mode shown in Fig. \ref{fig10}d, we find $\frac{2\rho_s}{|n|}=780\text{meV}\times a_0^2$ for a carrier density of $|n|=6\times 10^{11} \frac{1}{\text{cm}^2}$ per valley, resulting in a stiffness of $\rho_s = 280\mu \text{eV} = 3.28$K. In the realistic  system, the spins in the two valleys are coupled via a Hund's coupling, leaving only one gapless magnon mode (instead of one for each valley in our calculation, which neglects the inter-valley Hund's coupling). Assuming that the Hund's coupling is small, the total spin stiffness is then $\rho_{s,tot} = 560\mu \text{eV} = 6.56$K.

\section{Linearized gap equation}
Following Eq. \eqref{eq:coulomb_matrix_element}, we use
a shorthand notation for the intervalley interaction: 
\begin{equation}
    V_{+-}(\bm{k}_1,\bm{k}_2;\bm{k}'_1,\bm{k}'_2)= \braket{(+,\bm{k}_1),(-,\bm{k}_2)|v|(+,\bm{k}'_1),(-,\bm{k}'_2)}
\end{equation}
The interaction in the zero center of mass momentum pairing channel induced by exchange of a single particle-hole pair with valley charge can be written as 

\begin{align}
    V_\text{IVC} (k,k')=& \int \frac{d\bm{k}_1}{(2\pi)^2} \frac{d\bm{k}_2}{(2\pi)^2} \Big[ V_{+-}(\bm{k}_1+\bm{q},-\bm{k};\bm{k}',\bm{k}_1) 
    \chi\left[(+,\bm{k}_2+\bm{q}),(-,\bm{k}_2)|(+,\bm{k}_1+\bm{q}),(-,\bm{k}_1));\Omega_{\bm{q}}\right] 
    \nonumber \\ & \times V_{+-}( \bm{k},\bm{k}_2;\bm{k}_2+\bm{q},-\bm{k}') \Big],
\end{align}
where we use $k=(\bm{k},\Omega)$ as a shorthand notation, and $(\bm{q},\Omega_{\bm{q}})=k+k'$. 
It is worth noting that for a contact interaction, $V_{+-}(\bm{k}_1,\bm{k}_2;\bm{k}'_1,\bm{k}'_2)=U_{\pm}$, $V_\text{IVC}$ is simplified to $V_\text{IVC} (k,k')=U_{\pm}^2\tilde{\chi}_{\tau_+}(\bm{q},\Omega_{\bm{q}})$. Thus, $\chi_{\tau_+,\omega=0}$ 
can serve as a proxy for $V_\text{IVC}$. That being said, in our calculations we include the full momentum dependence of $V_\text{IVC}$ projected to the active band. 

We can write the vertex equation as 
\begin{equation}
    \Delta_{\bm{k}}(\Omega)=\int \frac{dk'}{(2\pi)^3} G(+,k')G(-,-k') \left[ V_{+-}(\bm{k},-\bm{k};\bm{k}',-\bm{k}')+ V_\text{IVC} (k,k')\right] \Delta_{\bm{k}'}(\Omega'),
    \label{eq:full_gap}
\end{equation}
where $G(\pm,k)=\braket{c_{\pm,k}^\dagger c_{\pm,k}}$ is the quasi-particle propagator.

As usual in problems of pairing mediated by a soft bosonic mode, there are several important energy scales in the problem: the Fermi energy $E_\mathrm{F}$, and the characteristic frequency $\Omega_{\text{IVC}}$ of the soft IVC mode (which we can identify as the frequency of the peak in the imaginary part of the IVC susceptibility, shown in Fig. \ref{fig3}d of the main text). $\Omega_{\text{IVC}}$ sets the scale for the frequency dependence of the effective interaction $V_\text{IVC}$: at frequencies below $\Omega_{\text{IVC}}$, $V_\text{IVC}$ becomes essentially frequency-independent. Note, however, that right at the critical point, $\Omega_{IVC}\rightarrow 0$. 

Solving the full frequency-dependent problem, Eq. \eqref{eq:full_gap}, is complicated, and requires (in addition to computing the full frequency dependence of $V_{\text{IVC}}$) also knowing the fully dressed fermionic Green's functions, including the self-energy due to the scattering of quasi-particles off IVC fluctuations. 
This problem has not been fully solved; nevertheless, there are both analytical and numerical indications that superconductivity is strongly enhanced near the quantum critical point~\cite{abanov2001coherent,Gerlach2017,lederer2017superconductivity}. 
Solving this problem for critical IVC fluctuations is beyond the scope of the current work. Here, we instead employ a crude approximation, ignoring the frequency dependence of $V_{\text{IVC}}$ and using the HF fermionic Green's functions.   
This yields the BCS-like gap equation discussed in the main text:
\begin{equation}
    \Delta_{\bm{k}}=\int \frac{dk'}{(2\pi)^3} G(+,k')G(-,-k') \left[ V_{+-}(\bm{k},-\bm{k};\bm{k}',-\bm{k}')+ V_\text{IVC} (\bm{k},\bm{k}')\right] \Delta_{\bm{k}'}.
    \label{eq:BCS_appendix}
\end{equation}

\begin{figure}
    \centering
    \includegraphics[width=0.7\linewidth]{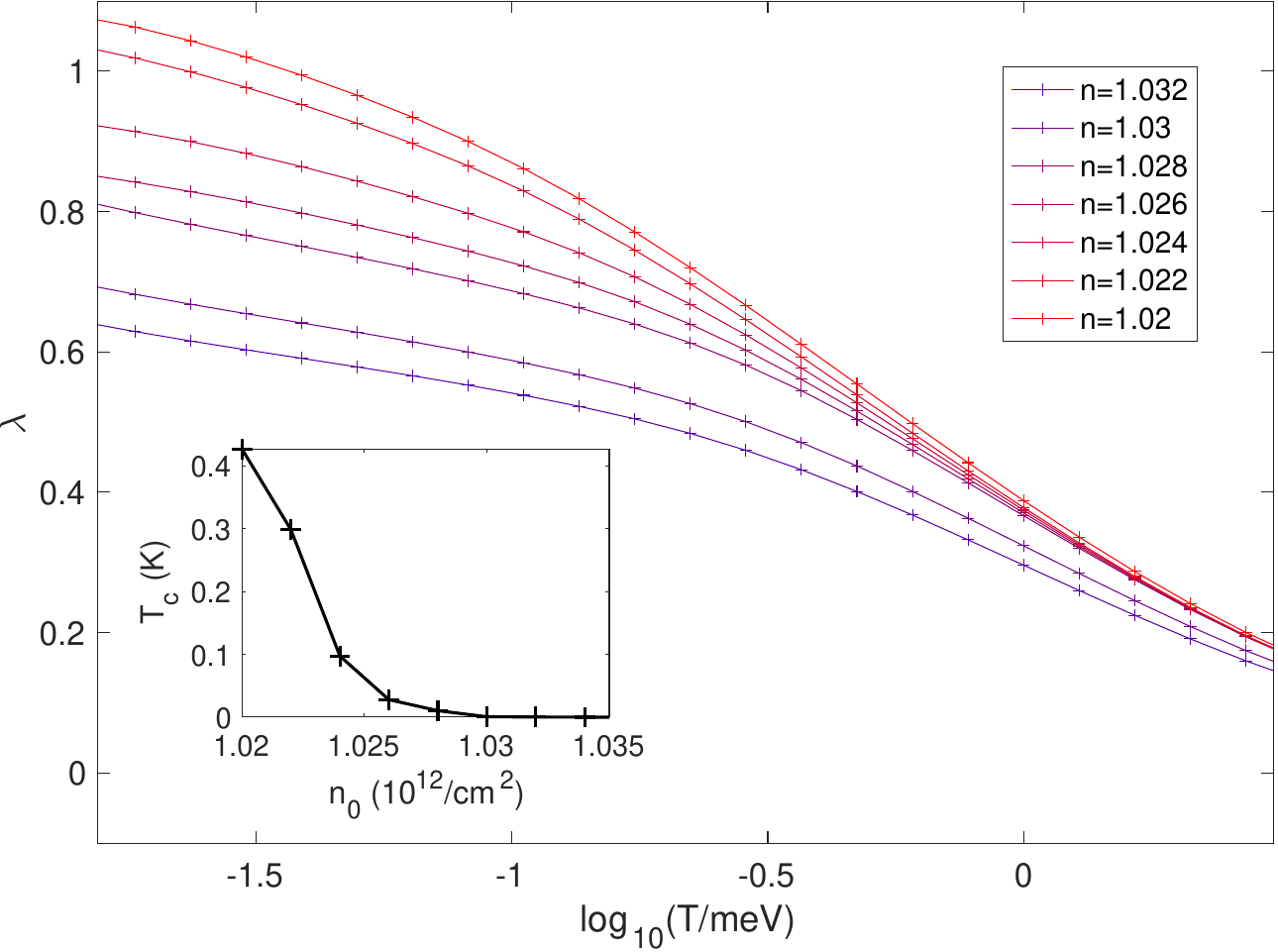}
    \caption{Largest eigenvalue, $\lambda$, of the linearized gap equation \eqref{eq:BCS_appendix} as function of temperature for different densities. Inset: critical temperature defined by $\lambda(T_c)=1$. 
    For some of the data points, $T_c$ is extracted by linearly extrapolating $\lambda$ vs. ${\rm log}(T)$ from the low temperature part of the data.}
    \label{fig:critical_temp}
\end{figure}

As mentioned in the main text, the effects of retardation of the interaction and the singular fermion self-energy are likely to decrease $T_c$ compared to the estimates from Eq. \eqref{eq:BCS_appendix}. Nevertheless, we expect our simple approximation to give the correct order of magnitude and trends of $T_c$, as well as the correct symmetry of the order parameter.

In order to be able to solve the linearized gap equation at low temperatures we need a very high momentum grid resolution (namely, $v_\mathrm{F}\Delta k \lesssim T$). In practice, using reasonable computational resources, the generalized RPA method cannot be implemented with a sufficiently high resolution. We overcome this problem by extracting the HF spectrum (i.e. the quasi-particle propagator) and the effective interaction using the generalized RPA method at a given grid resolution ($\Delta k=0.004\frac{1}{a_0}$), and then interpolate both the spectrum and the effective pairing interaction to achieve a better grid resolution ($\Delta k=0.0016\frac{1}{a_0}$), on which we solve the linearized gap equation to get the results plotted in Fig.~\ref{fig:critical_temp}.

\end{widetext}
\end{document}